# Scalable Solvent-Based Fabrication of Thermo-Responsive Polymer Nanocomposites for Battery Safety Regulation

*Mingqian Li, [a] Panpan Xu,[a] Suk-woo Lee,[b] Bum-young Jung,[b] Zheng Chen [a,c,d,e]* *

*[a]Department of NanoEngineering, University of California San Diego, La Jolla, CA 92093*

*[b] LG Energy Solution. Research Park, 188, Munji-ro, Yuseong-gu, Daejeon, 34122, Republic of Korea [c]Program of Chemical Engineering, University of California San Diego, La Jolla, CA 92093*

*[d]Program of Materials Science and Engineering, University of California San Diego, La Jolla, CA 92093*

*[e]Sustainable Power & Energy Center (SPEC), University of California San Diego, La Jolla, CA 92093*

*\*Correspondence to: zhengchen@eng.ucsd.edu*

## Abstract

Improving lithium-ion batteries (LIBs) safety remains in a challenging task when compared with the tremendous progress made in their performance in recent years. Embedding thermo-responsive polymer switching materials (TRPS) into LIB cells has been proved to be a promising strategy to provide consistent thermal abuse protections at coin-cell level. However, it is unrealistic to achieve large-scale applications without further demonstration in high-capacity pouch cells. Here, we employed tungsten carbide (WC) as a novel conductive filler, and successfully overcame the intrinsic processing difficulty of polyethylene (PE) matrix in a scalable solvent-based method to obtain ultra-thin, uniform, highly conductive TRPS. Moreover, by integrating TRPS directly into LIB electrodes, no extra fabrication facilities or processes are required for making the cells. As a result, multi-layer pouch cells with consistent electrochemical performance and thermal abuse protection function were fabricated using industry relevant manufacturing technique, which brings TRPS one step further to the real application scenarios.

Keywords: Lithium-ion batteries, safety, thermo-responsive composites, current collector

## Introduction

Great advances have been made in lithium-ion batteries (LIBs) since they were first invented in 1970s. [1, 2] High energy density,[3, 4] fast charging rate [5, 6] and wide temperature operation [7, 8] have been achieved, which paves their way to broad applications in portable electronic devices, [9] electric vehicles (EVs)[10] and grid energy storage. However, despite their rapid performance improvement in recent years, safety issue is still the sword of Damocles of LIBs due to their inherent flammability [11] and thermodynamically non-equilibrium operation, [12] which can lead to catastrophic consequences once they are exposed to extreme conditions such as overheating, shorting or cracking.[13]

Strategies to foreclose potential safety hazards in LIBs have been explored in different ways. In the aspect of system-level development, combined with big data and machine learning,[14] battery management systems (BMSs) are designed to collect critical cell information such as voltage, current, resistance and temperature to predict and respond ahead of dangers.[15] In terms of structural engineering, new cell configurations,[16] pack and modules arrangement [17] have been explored to reinforce the strength of battery packs against mechanical abuses.[18] Air/liquid cooling systems are also introduced to alleviate heat accumulations. However, all the aforementioned technologies evade confronting with the intrinsic issues from the materials level. As a result of delayed response and feedback, these approaches cannot eliminate safety risks instantly.

Significant efforts have been also devoted to designing internal safety strategies at the materials level. Functional separators, [19, 20] electrolyte additives,[21] non-flammable electrolytes [22, 23] and all-solid-state electrolytes (ASSEs) [24] have been explored to improve intrinsic cell safety. They are often realized by performance compensation and additional cost for device components and fabrications. [25-27] Thermo-responsive polymer switching materials (TRPS), which are capable to effectively protect batteries from thermal abuses, [28, 29] can achieve a good balance between safety, performance and costs. However, it cannot be widely applied unless a low-cost and scalable method is developed to obtain quality-consistent TRPS materials without interfering with the industry battery fabrication process.

In this work, we developed an efficient and scalable method to fabricate polyethylene (PE)-based ultra-thin, uniform, highly conductive TRPS for potential large-scale applications in LIBs. Using commercially available tungsten carbide (WC) nanopowder as the conductive filler, we elucidated that the embedded conductive network, which determines the performance of TRPS, largely depends on the size distribution of the conductive fillers and the fabrication process. Systematic studies on fillers and polymer selection, slurry preparation and film manufacturing were conducted to illustrate the correlation between fabrication process, filler distribution and TRPS properties. TRPS films were assembled under various conditions and inspected by surface and bulk characterization to exhibit filler distribution in the polymer matrix. Furthermore, electrical properties and thermal-responsive properties were measured. Successful assembly of TRPS-integrated cathodes can protect pouch cells from thermal abuse conditions, which paves the way for scalable manufacturing and application of TRPS in LIBs.

**Conductive filler selection**. To ensure that the TRPS films can function well, a good conductive filler should reach the following properties: (1) high intrinsic electrical conductivity at room temperature (R.T.), (2) good electrochemical stability in the battery operation window, and (3) small thermal expansion coefficient in a wide temperature range (R.T. to 200 $^{\circ}$C). Based on the above requirements, WC stands out as a promising candidate. WC is one of the most conductive materials among the carbide family, which has a bulk electrical conductivity $\sim 5.2 \times 10^4$ S cm$^{-1}$.[30] As a comparison, the conductivity of typical cathode/anode is in the range of 0.01 to 5 S cm$^{-1}$.[31-34] In addition, the electrochemical stability of WC was measured by cyclic voltammetry for both cathode and anode sides with coin cells embedded with polyethylene-based TRPS containing 80 wt.% of WC (PE-80WC). The result indicates that WC is able to tolerate wide electrochemical window (**Figure S1**) for LIB applications. In addition, WC has one of the smallest thermal expansion coefficients in common carbides/metals (**Table S1**). For a better control and simplicity of the fabrication, here we use commercial WC nanopowder (Alfa Aesar, 99.5%, **Figure S2**) as the conductive fillers to fabricate the TPRS and explore the structure-property relationship.

**Table 1. Physical properties of polymer candidates for TRPS.**

| | Thermal expansion coefficient -α ($10^{-6}$ m/(m•°C)) | Melting temperature $T_m$ (°C) | Hansen solubility parameter | | | Interaction radius | RED |
|---|---|---|---|---|---|---|---|
| | | | $\delta_d$ | $\delta_p$ | $\delta_h$ | | |
| LP40 | / | / | 17.3 | 14 | 4.3 | / | / |
| Polyethylene | 200 | 99-138 | 16.9 | 0.8 | 2.8 | 6.6 | 2 |
| Poly(vinylidene fluoride) | 125 | 158-200 | 19.4 | 15.9 | 11.3 | 9.6 | 0.9 |
| Polypropylene | 105 | 120-176 | 17.7 | 2.9 | 1.2 | 6.2 | 1.9 |
| Ethylene-vinyl acetate | 205 | 58-112 | 18.9 | -1.8 | 3.1 | 7.6 | 2.1 |
| Cellulose acetate butyrate | 145 | 150-240 | 17.2 | 13.8 | 2.8 | 12.6 | 0.1 |
| Polytetrafluorethylene | 165 | 317-345 | 16.2 | 1.8 | 3.4 | 3.9 | 3.2 |

**Polymer selection.** To exhibit an effective thermal response, a qualified polymer matrix must also satisfy several requirements: (1) large thermal expansion coefficient ($\alpha$), (2) suitable melting temperature ($T_m$), (3) electrochemically and physically compatible with electrolytes in the operation process, and (4) wide availability and low cost. Considering the aforementioned standards, PE, is the best candidate among promising candidates such as polyvinylidene fluoride (PVDF), polypropylene (PP), ethylene-vinyl acetate (EVA), cellulose acetate butyrate (CAB) and polytetrafluorethylene (PTFE). As one of the most commonly used polymers, it has a large $\alpha$ around $200 \times 10^{-6}$ K$^{-1}$ with a $T_m$ in the range of 100 ~ 140 °C, which enables the largest thermal expansion response of TRPS near the typical self-heating temperature (the onset of the battery self-heat process, i.e., 115.2 °C for the LiNi$_{0.5}$Co$_{0.2}$Mn$_{0.3}$O$_2$ (NCM523)/graphite pouch cells) of LIBs before thermal runaway.[35] Composing only carbon and hydrogen atoms, PE is employed as the separator material for LIBs. Due to the demonstrated stability, there would be no chemical interaction when PE is used as the polymer matrix for TRPS. However, to minimize the viscosity and maximize the thermal expansion, low-density PE is preferred. Thus, its solvent compatibility, such as swelling effect, should be evaluated.

Swelling is the volume expansion of polymer due to the penetration of solvent molecules, which can deteriorate the switching capability. Hansen solubility parameters (HSB) [36] are introduced to quantitatively describe the relative energy difference (RED) between the solvent and polymer, which are:

$$RED = \frac{R_a}{R_0} \tag{1}$$

$$(R_a)^2 = 4(\delta_{d2} - \delta_{d1})^2 + (\delta_{p2} - \delta_{p1})^2 + (\delta_{h2} - \delta_{h1})^2 \tag{2}$$

where $R_a$ is the Euclidean distance between the HSP for the solvent (s) and polymer, and it can be calculated by 3 solubility parameters: (1) $\delta_d$, dispersion forces, (2) $\delta_p$, polarity, and (3) $\delta_h$, hydrogen bonding; $R_0$ is the interaction radius. To estimate the physical compatibility of polymer in the electrolytes, ethylene carbonate (EC)/ diethyl carbonate (DEC) (50/50 vol.%) was selected as the target system to evaluate the polymer-solvent interaction. The results are presented in **Table 1**. $RED < 1$ indicates that the polymer can swell and dissolve in the solvent. While PE has a $RED = 2$, so it is an excellent anti-swelling polymer against common electrolytes. The overall properties of different polymer candidates are summarized in **Figure 1**, which confirms that PE is the best candidate as a polymer matrix for TRPS.

**Slurry preparation and film casting**

Common PE films are fabricated by film extrusion, injection molding and rotational molding, [37, 38] which require bulky and expensive equipment. To fabricate PE-based TRPS films with a small thickness (i.e., ~ 10 μm or lower), a low-cost and convenient solvent-based casting method was developed in this work. Again, HSB model was employed to find a suitable solvent. Several non-polar solvents such as *p*-xylene, limonene, *n*-butylbenzene and cyclohexane are potential candidates since their $RED$ with PE is smaller than 1 (**Table S2**). On the other hand, viscosity and volatility are critical in the film fabrication process. As a result, *p*-xylene was selected as the solvent for this study due to its good solubility, small viscosity and reasonable volatility. A heat-insulating vessel was designed for the Thinky mixer to obtain consistent and uniform mixing conditions. A typical solvent-cast fabrication procedure is shown in **Figure 2**. There are three major steps: mixing, casting and drying. First, PE, WC and xylene solvent were added into the container with desired ratios, then the container was transferred into a convection oven (80 °C) so that PE could be fully dissolved into the xylene solvent. After that, the precursor was thoroughly mixed in a Thinky mixer within the vessel to maintain the dissolving temperature of PE. Finally, the slurry was cast onto a glass/current collector in an pre-heated oven (at 80 °C) and dried out to get the film.

The spatial distribution of fillers in the polymer matrix, which determines the TRPS properties, highly correlates to the fabrication process. Depending on the drying temperature, evaporation rate and slurry viscosity, fillers will continually rearrange their distribution before their position being fixed. To demonstrate filler reorganization during the drying process, time-based experiments were designed (**Figure 3a**): a culture dish was used to limit the evaporation rate by isolating the air convection from the slurry surface and therefore controlling the "soak time", which indicates the duration of filler rearrange process. The cross-sectional scanning electron microscopic (SEM) images of two TRPS films with 60 wt.% of WC fillers exhibit distinct distribution behaviors upon different soak time. A uniform filler distribution with consistent conductivity (within a maximum of 1 order of magnitude difference) on both sides was achieved when the film dries out instantly after casting (**Figure 4a**), while an extended drying procedure leads to a filler separation and rearrangement, resulting in a 9-order of magnitude of conductivity difference between the top and bottom surface of the TRPS (**Figure 3b**).

Moreover, the size distribution of WC is also a decisive parameter of TRPS properties. **Figure 3c-d** show the conductivity and thermal switching capability of TRPS films fabricated with WC fillers with different particle size range, which were prepared by screening through various mesh sizes (1000 and 500 mesh, Figure S3). TRPS films integrated with pristine WC particles (P-TRPS) have the highest conductivity but the slowest switching response under a high temperature (100 °C). Further investigation from the SEM images in **Figure S4** shows that big WC chunks exist in P-TRPS. Large particles can facilitate the electron transport through the film more effectively but can also attenuate switching performance since it transfers electrons by bulk instead of inter-particle network. For the following demonstration, all WC particles were pre-treated by 1000 mesh screening and the TRPS films were prepared by immediate drying (soak time = 0 min) after casting to maintain a uniform spatial distribution of WC particles.

**Basic properties of TRPS**

TRPS films were prepared and the surface conductivity on both sides as well as the vertical resistance were measured as a function of WC contents. As shown in **Figure 4a**, the surface conductivity reaches around 2 S cm$^{-1}$ from 10$^{-8}$ S cm$^{-1}$ as the WC weight ratio increases from 65% to 85%, with a conductive threshold of 60 wt.%. There is 1 order of magnitude difference in terms of conductivity between the top and bottom surface at low WC contents (below 75 wt.%), which might be attributed to the non-uniform distribution of WC in the PE matrix. However, the conductivities become consistent as WC content increases due to the formation of a robust conductive network. On the other hand, the vertical resistance rapidly reaches the same order as the robust network forms, and the resistance slightly decreases to ~ 3 Ω. Such anisotropic conductivity property of the as-prepared TRPS may provide more possibilities for other applications such as electronic sensors. [39, 40]

To investigate the potential of the as-prepared TRPS films for battery safety application, switching capability was measured within a coin cell-like setup (**Figure S5**) (**Figure 4b**). TRPS with WC ratio between 80 wt.% to 85 wt.% were selected and stored in a 100 °C convection oven to stimulate a thermal-abuse condition. The PTC intensity (the logarithm of the ratio of the maximum resistivity to the resistivity at room temperature) increases to 3 for WC/PE films with 80 wt.%. The TRPS barely shows switching response if continues raising WC contents to 85 wt.% (PTC = 0.3). Considering both the PTC intensity and the R.T. conductivity, TRPS films with 80 wt.% seems to be a good composition for real battery applications.

To further exam the switching response rate corresponding to various thermal-abuse conditions in different scenarios, TRPS films were integrated into both coin-cell and pouch-cell structures to be tested using a hot-air gun under different heating temperatures (**Figure S6**). As shown in **Figure 4c**, a higher heating temperature (250 °C) gave rise to a more rapid and intense response of resistance change than a lower temperature (150 °C). Compared with coin-cell architecture, TRPS embedded in a pouch structure reveals a faster and more intense response to temperature changes, which indicates that the PTC intensity as well as the switching response rate are highly correlated to the cell configuration and thermal abuse conditions. Overall, an instant and rapid temperature change will induce a more prompt and significant resistance increase. More details with temperature and resistance change with time were tracked and presented in **Figure 4d**. The rapid increase of temperature from R.T. to 110 °C triggers rapid increase of resistance without any delay (**Figure 4d, inset**), which may ensure a real time shutdown of battery at safety events to prevent catastrophic consequences. In addition, the maximum resistance is about 10 times of that obtained at low heating temperature (150 $^{o}$C, in a pouch-like structure), which further ensures the capability of the TRPS to immediately shut down cells at an extreme heating rate or high current conditions.

Ideally, TRPS films should maintain a high resistance under extreme temperature conditions (e.g., aging at 100 $^{o}$C) for extended operation time while exhibiting stable resistance under normal temperature for battery operation, which are further demonstrated. As show in **Figure 4e**, even though the resistance fluctuates over time (1500s), the resistance value is higher than that obtained at the shutdown condition (**Figure 5d**). Meanwhile, the TRPS films are capable of retaining small resistance over a wide temperature range to satisfy the wide operation window of LIBs. In **Figure 4f**, the resistance change is about ~0.001 $\Omega$ s$^{-1}$ (0.065% of its initial resistance (1.54 $\Omega$)) below 45 $^{o}$C, even when the temperature goes up to 70 $^{o}$C, the resistance change rate is only ~ 0.02 $\Omega$ s$^{-1}$, which confirms that TRPS will not limit the normal operation of LIBs at the relatively high-temperature scenario.

### LIB cell thermal abuse testing

To verify the safety properties of the TPRS for LIB applications, LiNi$_{1/3}$Co$_{1/3}$Mn$_{1/3}$O$_2$ (NCM111)/graphite pouch cells were fabricated with a temperature sensor embedded inside to monitor the temperature variations (**Figure 5a**). The normal Al current collector was replaced with the TRPS- coated Al (Al-TRPS, inset of **Figure 5b**) when assembling the cell. A TRPS layer with a thickness of around 20 µm is in intimate contact between the Al substrate and the NCM active layer, which was prepared through the aforementioned solvent casting method. The cycling results of the TRPS-integrated pouch cell and the control pouch (TRPS-free) are compared in **Figure 5b**, which shows similar capacity and stability at 25$^{o}$C.

Thermal abuse conditions were simulated by exposing the cycling TRPS-pouch to a convection oven at 100 $^{o}$C to investigate the response of TRPS-cells against temperature variation. As shown in **Figure 5c**, at the beginning the TRPS-pouch cycled very well at R.T. conditions and then shut down immediately when it suffered from an overheating at 100 $^{o}$C in the oven. Meanwhile, it can resume to normal charge/discharge cycling once the temperature returned to the R.T. In thermal abuse conditions, the TRPS-pouch can effectively shut down for multi-times at various C-rates or charge/discharge conditions (**Figure S7a-b**). To further verify the TRPS function in a more realistic condition, a bi-layer pouch with a capacity of 70 mAh was fabricated and tested (**Figure S7c**), which shows the same switching result. This experiment indicates that such an Al-TRPS-NCM integrated electrode can be extended to real multilayer pouches with excellent thermal switching capacities for improved LIB safety.

In addition, the evolution of resistance and temperature were recorded during a simulated overheating process (**Figure 5d**). The cell resistance reached to above 600 $\Omega$ after being exposed to the convection oven at 70 $^{\circ}$C for 20 s, showing a rapid and intense total thermal response. As the cell was intentionally cut off at 4.29V in our experiment settings, the actual maximum cell resistance upon switching should be significantly larger than the calculated value based on Ohm's law. Nevertheless, the resistance curve is synchronous with the temperature curve, which indicates a fast and sensitive cell resistance change upon heating the TRPS in the pouch.

**Cell electrical abuse testing**

To more comprehensively inspect the safety protection capabilities of TRPS, electrical abuse tests were also conducted, and the results are shown in **Figure 6a**. To simulate the external shorting, a large current at 120 C rate (corresponding to ~200 mA, with a voltage upper limit at 21 V) was forced to pass through the cell, which can generate a large amount (2.2 W) of heat to trigger the thermal response of the TRPS and result in temperature ramping. As a result of the thermal response, the temperature of the TRPS cell stopped increasing at around 45 to 50 $^{\circ}$C, while the temperature of the control cell continued to ramp to ~90 $^{\circ}$C until the end of the simulated external short test. The results from resistance and temperature response with time indicate a rapid resistance change (PTC ~ 2.1) occurred after the initial temperature ramping due to the large current which leads to the reduction of "short" current. This greatly reduces the joule heating and therefore prohibits further temperature increase of the cell. On the contrary, the cell without TRPS showed negligible increase of the resistance and the temperature showed no sign of decrease up to 90 $^{\circ}$C before the end of the testing. (**Figure S8**).

**Conclusion**

In summary, a scalable solvent-based fabrication process of TRPS films from slurry preparation to casting was designed and optimized specifically for PE matrix. WC was employed as conductive fillers to assemble highly conductive thermo-responsive switching composites. Systematic studies on fillers and polymer selection were conducted to illustrate the essential requirements for a good filler-polymer matrix system for TRPS materials. The fabrication process and the size distribution of fillers were investigated extensively to demonstrate their decisive roles in the properties of TRPS. By integrating a thin TRPS directly into the current collector, LIB pouch cells with thermal abuse protection features can be assembled without additional modification of current industrial fabrication processes, which brings TRPS one step closer to large scale applications in LIBs.

### Experimental section

*TRPS film fabrication*: WC (Alfa Aesar, 99.5%) were dispersed into isopropyl alcohol (Sigma-Aldrich) and filtered through a 1000-mesh screen followed by drying out in a convection oven (MTI EQ-DHG-9015) at 55 °C for 2 hours. Then the pretreated WC was mixed with PE (Epolene C-13 $M_w$ =76000) at different ratio in xylene by a Thinky mixer (Model ARE-310) in a designed heat-insulating container. The slurry was then coated on a glass substrate or carbon-coated Al current collector (Al-TRPS) with vacuum drawing and by a doctor blade. The slurry was dried out immediately. The slurry was also tested with different soaking time to study the effect of particle distribution and uniformity on the properties of the obtained TRPS films.

*TRPS film properties testing:* For conductivity tests, the surface electrical conductivity at R. T. was measured by a four-point probe with a Keithley 2400 and the vertical resistance at R.T. was measured by a two-point probe. For the intensity of switching response (PTC intensity), the time-dependent resistance measurements of various composites were conducted with a two-probe method by putting samples into a convection oven at 100 °C with setups shown in **Figure S6a-b**. The resistance was recorded by a Keithley 2400. The time-dependent temperature measurements of TRPS were tracked by a K-type thermal couple on a temperature logger. For rate of switching response, a hot-air gun (YiHua SMD Reworks Station 852D+) was used for response rate testes by heating samples under different temperatures (150 °C and 250 °C) for various cell configurations. The resistance and temperature changes were recorded by the same way as switching resistance response tests.

*Battery fabrication*: LIB cathode were made by mixing NCM111, polyvinylidene fluoride (PVDF) and Super P at a mass ratio of 8:1:1 in N-Methyl-2-pyrrolidone (NMP) solvent and then coated and dried in vacuum at 80 °C for 10 h. Anodes were made by mixing graphite, PVDF and Super P at a mass ratio of 90:5:5 in NMP solvent and then coated and dried in vacuum at 80 °C for 10 h. The electrolyte was 1 M $LiPF_6$ in ethylene carbonate/diethylcarbonate (1:1 v/v) (LP40). Pouch cells were fabricated with areal capacity of 1.5 mAh $g^{-1}$ and N/P ratio at ~1.1-1.2. The cathode size was 57mm × 44 mm, and the anode size was 58 mm × 45 mm for pouch cells. The typical capacity for a single-layer pouch was about 30 mAh. Al-TRPS-NCM cathodes were prepared by directly cast cathode slurry on the Al-TRPS substrates, and TRPS-pouch cells were fabricated the same method as the control cells.

*Battery testing:* Galvanostatic charge/discharge cycling was performed on Neware BTS-4000 for tracking cycling performance. For thermal abuse testing, the TRPS-cells were tested in a convection oven (MTI EQ-DHG-9015) held at 100 °C to examine the thermal shutdown capability. Specifically, the cells will be transferred into the convection oven (100 °C) when they were charging/discharging by the Neware BTS-4000 tester, and the voltage and temperature changes were recorded by theNeware software. To measure the real time temperature of cells, a K-type thermal couple was attached on the surface of pouch cells/coin cells, which is connected with a temperature logger. The temperature changed will be read and recorded by the HOBOware reader.. The external shorting tests were conducted with the setup shown in **Figure S5b** by applying an extra-large current (120C, corresponding to ~200 mA for the coin cells). The resistance and temperature changes were recorded by a Keithley 2400 and a K-type thermal, respectively.

*Acknowledgements*

This study was partially supported by the LG Chem through Battery Innovation Contest (BIC) program. The authors acknowledge the UCSD Crystallography Facility. A portion of the work used the UCSD-MTI Battery Fabrication Facility and the UCSD-Arbin Battery Testing Facility. The authors also acknowledge the start-up fund support from the Jacob School of Engineering at UC San Diego.

*Conflict of Interest*

A patent was filed for this work through the UCSD Office of Innovation and Commercialization.

# Reference

1. M. S. Whittingham, *J. Electrochemical Soc.*, **123** (1976).
2. M. S. Whittingham, *Science*, **192** (1976).
3. R. Weber, M. Genovese, A. J. Louli, S. Hames, C. Martin, I. G. Hill and J. R. Dahn, *Nat. Energy*, **4**, 683 (2019).
4. Z. Wang, J. Shen, J. Liu, X. Xu, Z. Liu, R. Hu, L. Yang, Y. Feng, J. Liu, Z. Shi, L. Ouyang, Y. Yu and M. Zhu, *Adv. Mater.*, **31**, 1902228 (2019).
5. H. Liu, Z. Zhu, Q. Yan, S. Yu, X. He, Y. Chen, R. Zhang, L. Ma, T. Liu, M. Li, R. Lin, Y. Chen, Y. Li, X. Xing, Y. Choi, L. Gao, H. S. Cho, K. An, J. Feng, R. Kostecki, K. Amine, T. Wu, J. Lu, H. L. Xin, S. P. Ong and P. Liu, *Nature*, **585**, 63 (2020).
6. S. X. H. Jin, C. Chuang, W. Li, H. Wang, J. Zhu, H. Xie, T. Zhang, Y. Wan, Z. Qi, W. Yan, Y. R. Lu, T. S. Chan, X. Wu, J. B. Goodenough, H. Ji and X. Duan, *Science*, **370**, 192 (2020).
7. J. Holoubek, Y. Yin, M. Li, M. Yu, Y. S. Meng, P. Liu and Z. Chen, *Angew. Chem. Int. Ed.*, **58**, 18892 (2019).
8. G. Cai, J. Holoubek, D. Xia, M. Li, Y. Yin, X. Xing, P. Liu and Z. Chen, *Chem. Commun. (Cambridge, U. K.)*, **56**, 9114 (2020).
9. Y. Liang, C. Z. Zhao, H. Yuan, Y. Chen, W. Zhang, J. Q. Huang, D. Yu, Y. Liu, M. M. Titirici, Y. L. Chueh, H. Yu and Q. Zhang, *InfoMat*, **1**, 6 (2019).
10. G. Zubi, R. Dufo-López, M. Carvalho and G. Pasaoglu, *Renew. Sust. Energ. Rev.*, **89**, 292 (2018).
11. C. Arbizzani, G. Gabrielli and M. Mastragostino, *J. Power Sources*, **196**, 4801 (2011).
12. S. Abada, G. Marlair, A. Lecocq, M. Petit, V. Sauvant-Moynot and F. Huet, *J. Power Sources*, **306**, 178 (2016).
13. M. Ouyang., X. Feng, X. Liu, L. Lu, Y. Xia and X. He, *Energy Storage Mater.*, **10**, 246 (2018).
14. W. Li, J. Zhu, Y. Xia, M. B. Gorji and T. Wierzbicki, *Joule*, **3**, 2703 (2019).
15. R. Xiong, L. Li and J. Tian, *J. Power Sources*, **405**, 18 (2018).
16. D. L. Thompson, J. M. Hartley, S. M. Lambert, M. Shiref, G. D. J. Harper, E. Kendrick, P. Anderson, K. S. Ryder, L. Gaines and A. P. Abbott, *Green Chem.*, **22**, 7585 (2020).
17. D. Kang, P.-Y. Lee, K. Yoo and J. Kim, *J. Energy Storage*, **27** (2020).
18. A. A. H. Akinlabi and D. Solyali, Renewable and Sustainable Energy Reviews, **125** (2020).
19. M. S. Gonzalez, Q. Yan, J. Holoubek, Z. Wu, H. Zhou, N. Patterson, V. Petrova, H. Liu and P. Liu, *Adv. Mater.*, **32**, 1906836 (2020).
20. T. Lei, W. Chen, Y. Hu, W. Lv, X. Lv, Y. Yan, J. Huang, Y. Jiao, J. Chu, C. Yan, C. Wu, Q. Li, W. He and J. Xiong, *Adv. Energy Mater.*, **8** (2018).
21. B. R. R. T. Dagger, F. M. Schappacher and M. Winter, *Energy Technol.*, **6**, 1 (2018).
22. H. Sun, G. Zhu, Y. Zhu, M. C. Lin, H. Chen, Y. Y. Li, W. H. Hung, B. Zhou, X. Wang, Y. Bai, M. Gu, C. L. Huang, H. C. Tai, X. Xu, M. Angell, J. J. Shyue and H. Dai, *Adv. Mater.*, **32**, 2001741 (2020).
23. H. Yang, C. Guo, J. Chen, A. Naveed, J. Yang, Y. Nuli and J. Wang, *Angew. Chem., Int. Ed. Engl.*, **58**, 791 (2019).
24. T. Jiang, P. He, G. Wang, Y. Shen, C. W. Nan and L. Z. Fan, *Adv. Energy Mater.*, **10** (2020).
25. J. M. Doux, H. Nguyen, D. H. S. Tan, A. Banerjee, X. Wang, E. A. Wu, C. Jo, H. Yang and Y. S. Meng, *Adv. Energy Mater.*, **10** (2019).
26. J. M. Doux, Y. Yang, D. H. S. Tan, H. Nguyen, E. A. Wu, X. Wang, A. Banerjee and Y. S. Meng, *J. Mater. Chem. A*, **8**, 5049 (2020).

27. T. G. Joscha Schnell, Thomas Knoche, Christoph Vieider, Larissa Köhler, Alexander Just, Marlou Keller, Stefano Passerini, Gunther Reinhart, *J. Power Sources*, **382**, 160 (2018).

28. P. C. H. Zheng Chen, Jeffrey Lopez, Yuzhang Li, John W. F. To, Nan Liu, Chao Wang, Sean C. Andrews, Jia Liu, Yi Cui & Zhenan Bao, *Nat. Energy*, **1** (2016).

29. M. Li, Y. Shi, H. Gao and Z. Chen, *Adv. Funct. Mater.*, **30** (2020).

30. X. Guo, L. Li, Z. Liu, D. Yu, J. He, R. Liu, B. Xu, Y. Tian and H.-T. Wang, *J. Appl. Phys*, **104** (2008).

31. M. Ghalkhani, F. Bahiraei, G.-A. Nazri and M. Saif, *Electrochim. Acta*, **247**, 569 (2017).

32. J. Li, Y. Cheng, L. Ai, M. Jia, S. Du, B. Yin, S. Woo and H. Zhang, *J. Power Sources*, **293**, 993 (2015).

33. R. Zhao, J. Liu and J. Gu, *Appl. Energy*, **173**, 29 (2016).

34. Y. Wang, H. Zheng, Q. Qu, L. Zhang, V. S. Battaglia and H. Zheng, *Carbon*, **92**, 318 (2015).

35. X. Liu, D. Ren, H. Hsu, X. Feng, G.-L. Xu, M. Zhuang, H. Gao, L. Lu, X. Han, Z. Chu, J. Li, X. He, K. Amine and M. Ouyang, *Joule*, **2**, 2047 (2018).

36. C.M. Hansen, *Hansen Solubility Parameters A Users Handbook*, 2$^{nd}$ Ed ,ISBN-0849372488.

37. S.I. Hong and J.W. Rhim, LWT, *Food Sci. Technol.* ., **48**, 43 (2012).

38. F. Yu, H. Deng, H. Bai, Q. Zhang, K. Wang, F. Chen and Q. Fu, *ACS Appl. Mater. Interfaces*, **7**, 10178 (2015).

39. J. Yang, R. Downes, A. Schrand, J. G. Park, R. Liang and C. Xu, *Scr. Mater*, **124**, 21 (2016).

40. J. F. Gao, D.X. Yan, B. Yuan, H.-D. Huang and Z.M. Li, *Compos Sci Technol* ., **70**, 1973 (2010).

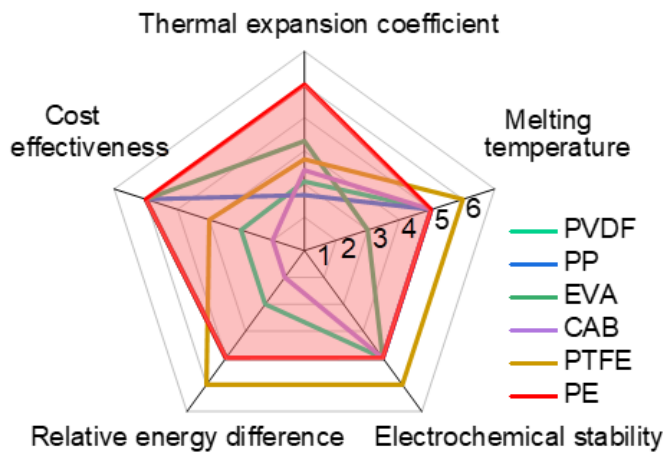

**Figure 1.** Comparison of physical properties of different polymer candidates as the potential TRPS matrix.

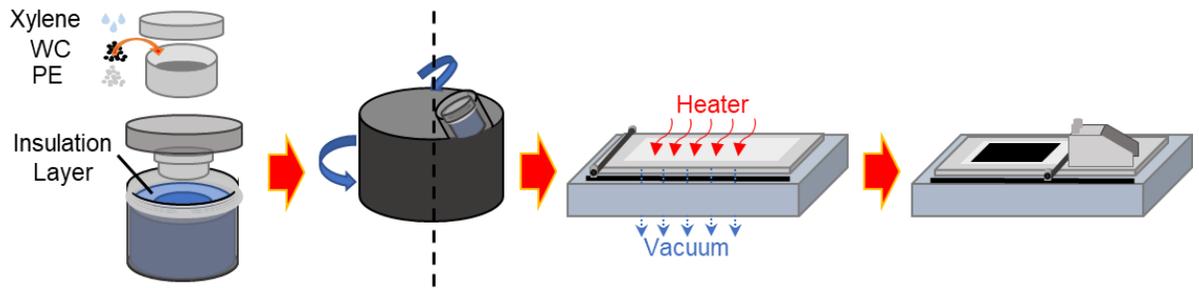

**Figure 2.** Schematic illustration of the solvent casting process to fabricated TRPS. Conductive fillers, solvent and polymer are mixed by a Thinky mixer within a heat-insulated container to form a homogenous slurry, which is cast by the auto-coater with a vacuum pump to fix the position.

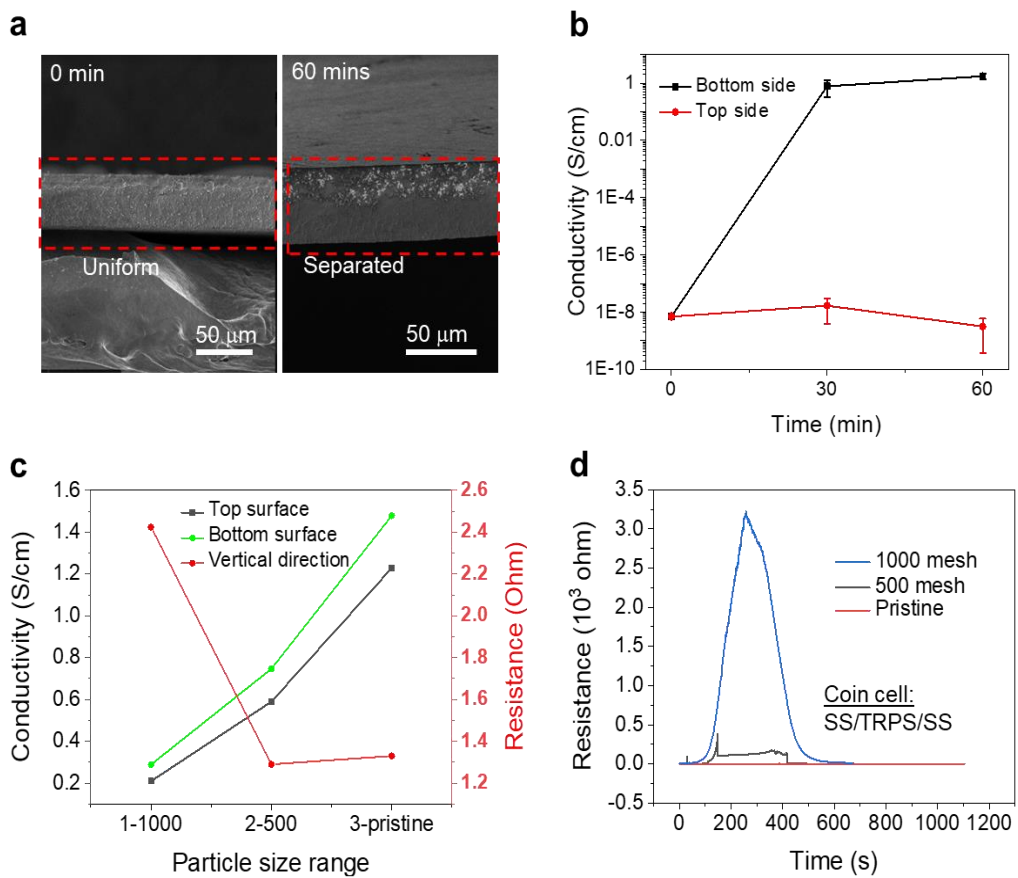

**Figure 3.** (a) Cross-section SEM images of TRPS samples prepared with different soaking time before drying (weight percent of WC was 60%; corresponding to a volume fraction of 16%). (b) Surface (lateral direction) conductivity of WC/PE films as a function of soaking time. (c) Top, bottom conductivities and resistance of the vertical direction of PE-80WC at different size ranges (soak time = 0 min). (d) Corresponding switching response of TRPSs made from WC with different average sizes.

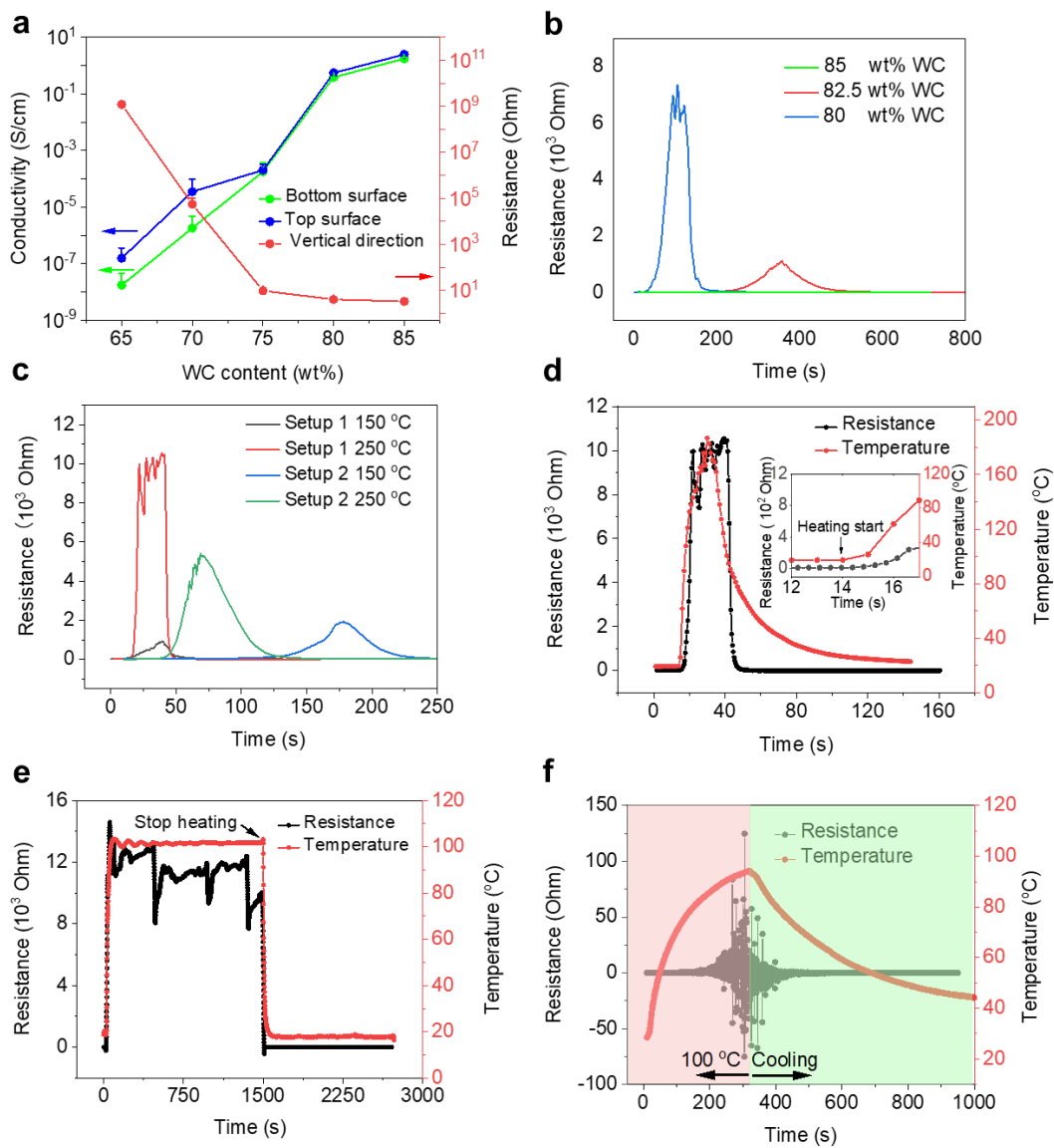

**Figure 4.** (a) Surface conductivity and vertical resistance of WC/PE films as a function of the WC ratio in the TRPS. (b) Thermal switching response of TRPS as a function of time for samples made with different WC ratios. The testing setups are shown in **Figure S4**.(c) Changes of temperature (surface) and resistance as a function of time for WC/PE films with 80 wt% of WC at different heating rates in a pouch cell. The setups are shown in **Figure S5**. (d) Changes of temperature (pouch surface) and resistance as a function of time for WC/PE film with 80% of WC at 250 °C heating rate in a pouch cell (e) Changes of resistance of WC/PE film in a convection oven held at 100 °C for 1500s. (f) Resistance variations per second and temperature change vs time of TRPS in the temperature chamber held at 100 °C.

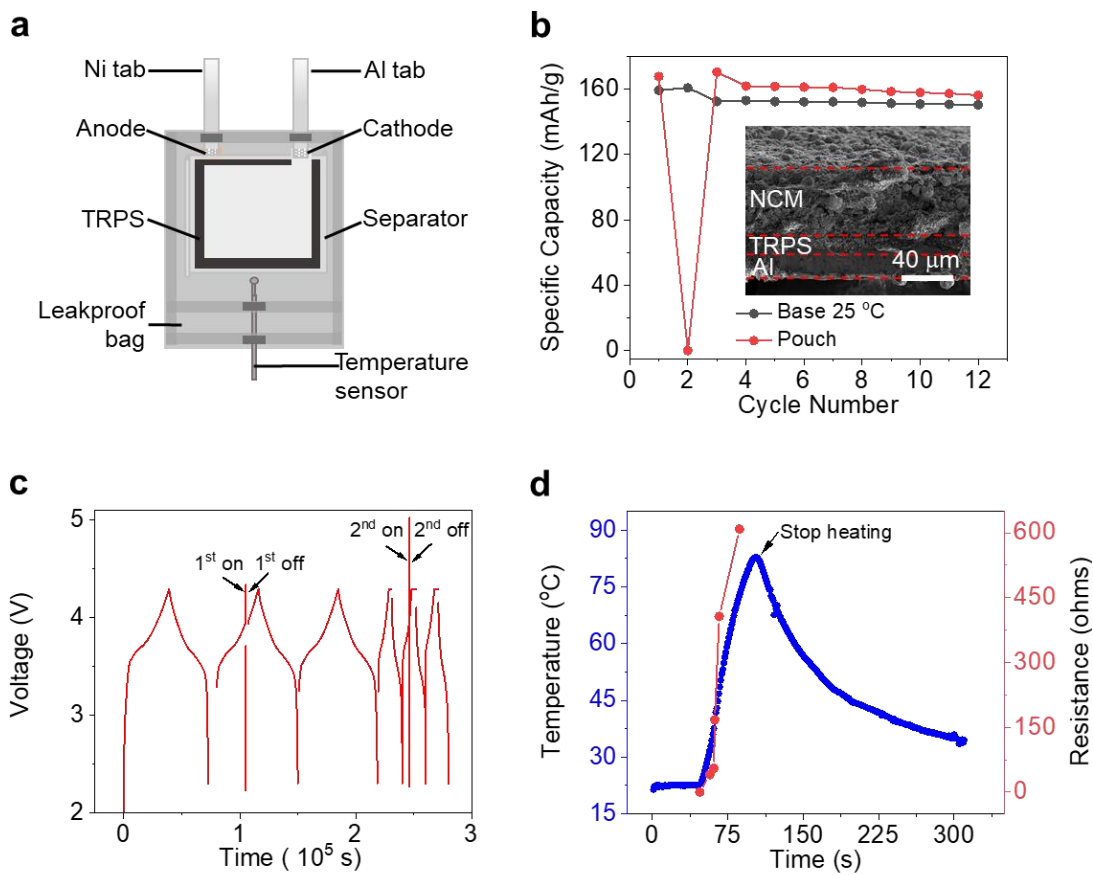

**Figure 5.** (a) Structure of the pouch cells for monitoring the temperature change *vs* time. The cathode size is 57 mm × 44 mm, and the anode is about 58 mm × 45 mm. The cathode mass loading is about 10 mg cm$^{-2}$ for all the cells, with a N/P ratio of ~1.1 to 1.2 for a full cell. (b) Cycling stability of the safety-improved TRPS pouch cell (NCM111/graphite) and a control cell without using any TRPS. Inset is the cross-sectional SEM image of a representative cathode made with NCM111 coated on an Al-TRPS current collector. (c) Charge-discharge curves of Al-TRPS-embedded NCM111/graphite pouches before and after thermal shut down in a 100 °C convection oven. (d) Resistance and temperature changes as a function of time for a NCM111/graphite pouch cell with Al-TRPS during and after a thermal abuse.

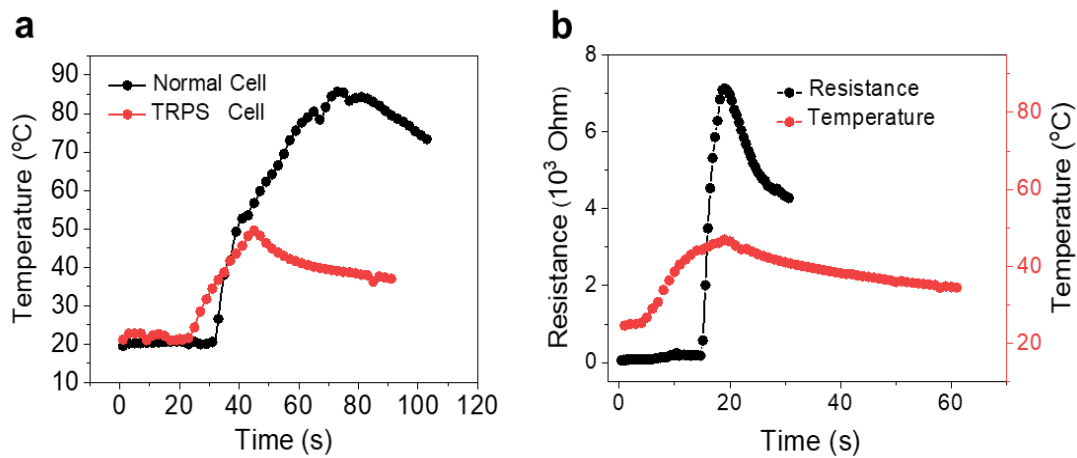

**Figure 6.** (a) Temperature profiles for different cells under external shorting using constant current (120 C). For the TRPS cell, the temperature increasing rate was 1.27 C s⁻¹, for the control cell the rate was 1.55 C s⁻¹. (b). Temperature and resistance as a function of time for the TRPS cell.